# High thermoelectric performance of half-Heusler compound BiBaK with intrinsically low lattice thermal conductivity

S. H. Han, Z. Z. Zhou, C. Y. Sheng, J. H. Liu, L. Wang, H. M. Yuan and H. J. Liu[*]

*Key Laboratory of Artificial Micro- and Nano-Structures of Ministry of Education and School of Physics and Technology, Wuhan University, Wuhan 430072, China*

Half-Heusler compounds usually exhibit relatively higher lattice thermal conductivity that is undesirable for thermoelectric applications. Here we demonstrate by first-principles calculations and Boltzmann transport theory that the BiBaK system is an exception, which has rather low thermal conductivity as evidenced by very small phonon group velocity and relaxation time. Detailed analysis indicates that the heavy Bi and Ba atoms form a cage-like structure, inside which the light K atom rattles with larger atomic displacement parameters. In combination with its good electronic transport properties, the BiBaK shows a maximum *n*-type *ZT* value of 1.9 at 900 K, which outperforms most half-Heusler thermoelectric materials.

## 1. Introduction

The search and development for new energy materials has become a top priority to overcome the increasingly serious energy crisis and environmental pollution. It is noteworthy that thermoelectric (TE) materials can convert heat into electricity directly, which has attracted widespread attention from the science community. The efficiency of a TE material depends on the dimensionless figure-of-merit (*ZT*), defined as $ZT = S^2\sigma T/(\kappa_l + \kappa_e)$, where $S$, $\sigma$, $T$, $\kappa_l$, and $\kappa_e$ are the Seebeck coefficient, the electrical conductivity, the absolute temperature, the lattice thermal conductivity, and the electronic thermal conductivity, respectively. Over the past two decades, many strategies have been implemented successfully aiming to increase the power factor ($S^2\sigma$) and/or decrease the $\kappa_l$ [1−6]. However, it remains a challenge to significantly enhance the thermoelectric performance due to the inherent coupling of $S$, $\sigma$, and $\kappa_e$ [7].

---

[*] Author to whom correspondence should be addressed. Electronic mail: phlhj@whu.edu.cn

Among various TE materials, half-Heusler (HH) compounds are potential candidates owing to their excellent electronic transport properties, which originate from moderate band gap and sharp density of states (DOS) around the Fermi level. For example, the $S^2\sigma T$ of HH alloys TiIrAs, ZrIrSb, and ZrCoSb are more than 6 W/mK at 800 K [8], which surpass those of many good thermoelectric materials. However, a majority of HHs exhibit higher lattice thermal conductivity in the order of magnitude of 10 W/mK, which seriously restricts the improvement of their ZT values [9−12]. During the past few years, more and more efforts have been focused on lowering the $\kappa_l$ to enhance the ZT values by nano-structuring, doping, and alloying [9,12−17]. For example, it was found that a maximum ZT of ~1.5 can been obtained at 1200 K for the p-type $FeNb_{1-x}Hf_xSb$ with $\kappa_l$ less than 5 W/mK [15]. Besides, a high ZT of ~1.2 can be achieved in the n-type $Zr_{1-x}Hf_xNiSn$ at 873 K, where the $\kappa_l$ is only about 2 W/mK [16]. However, these approaches may adversely affect the electronic transport properties. Hence, it is important to figure out what kind of HHs are prone to exhibit intrinsically lower $\kappa_l$.

Recently, it was reported that one of the HHs BiBaK exhibits unprecedentedly low $\kappa_l$ [18], which is calculated to be 2.19 W/mK at room temperature. It is thus interesting to check whether BiBaK could have good thermoelectric performance. In this work, we present a theoretical study on the structural, phonon, electronic and thermoelectric transport properties of BiBaK by adopting first-principles pseudopotential method and Boltzmann transport theory. We shall see that by optimizing the carrier concentration, a maximum ZT value of 1.9 can be achieved at 900 K in the n-type system, which is much higher than those found in most HHs in their pristine form.

**2. Computational method**

In order to obtain the phonon dispersion relations of half-Heusler compound BiBaK, we combine the density functional theory (DFT) [19] calculations with the finite displacement method. The former is implemented in the Vienna *ab-initio* simulation package (VASP) [20] and the latter in the PHONOPY code [21], where a $3\times3\times3$ and $4\times4\times4$ supercell are respectively used in the second- and third-order interatomic force constants (IFCs). To obtain the phonon transport properties, we solve the linearized phonon Boltzmann transport equation embedded in the so-called ShengBTE package [22]. During the calculations, a cutoff radius of 7.7 Å is imposed on the third-order interactions and we adopt a suitable *q*-mesh as large as $31\times31\times31$ to get converged lattice thermal conductivity.

Within the framework of DFT, the calculations of electronic properties of BiBaK are performed using the projector augmented wave (PAW) method, where the exchange-correlation energy is in the form of Perdew-Burke-Ernzerhof (PBE) under the generalized gradient approximation (GGA) [23]. The hybrid functional of Heyd-Scuseria-Ernzerhof (HSE) [24] is adopted for more accurate band gap and electronic transport coefficients with the effect of spin-orbit coupling (SOC) taken into account. We use a Monkhorst-Pack *k*-mesh of $15\times15\times15$ for sampling the Brillouin zone. The electronic transport coefficients including the Seebeck coefficient, the electrical conductivity, and the electronic thermal conductivity are evaluated from the semi-classical Boltzmann transport theory as implemented in the BoltzTraP code [25]. Considering the interaction between electrons and acoustic phonons, we use the deformation potential (DP) theory [26] to deal with the relaxation time.

## 3. Results and discussion

Ternary intermetallic HH compounds exhibit a crystal structure of $XYZ$ with space group $F\bar{4}3m$, where the $X$, $Y$, and $Z$ atoms are located in the Wyckoff positions of 4c ($\frac{1}{4}$, $\frac{1}{4}$, $\frac{1}{4}$), 4b ($\frac{1}{2}$, $\frac{1}{2}$, $\frac{1}{2}$), and 4a (0, 0, 0), respectively. It was suggested that the 8 or 18 valence electrons per primitive cell can predict the electronic properties of HHs [27−29], and their electronic structures are related to which of the three atoms occupies the 4c position [30]. In principle, one can have three inequivalent atomic configurations of HHs, namely, BiBaK, BaKBi, and KBiBa. Our first-principles calculations show that among them, the BiBaK exhibits the lowest energy with an optimized lattice constant of 8.45 Å. Besides, there is no imaginary frequency in the phonon spectrum, which means that the BiBaK compound is dynamically stable. To further test its stability, *ab-initio* molecular dynamics (AIMD) simulations have been performed. The AIMD runs for 5000 steps with a time step of 0.5 fs. Figure 1 shows the nearest Bi-Ba distance as a function of MD step at different temperatures. It can be seen that up to 900 K, there are only small fluctuations around the equilibrium bond lengths of 3.66 Å and the crystal structure remains unchanged. All these observations suggest that the BiBaK compound is rather stable.

As mentioned above, the BiBaK has ultralow lattice thermal conductivity compared with most HHs. To have a better understanding, we show in Figure 2 the phonon spectrums of BiBaK and other three HHs with $\kappa_l$ in excess of 10 W/mK at room temperature [31−33]. As the primitive cell of BiBaK contains three atoms, we see there are 9 phonon branches with 3 acoustic and 6 optical ones. The maximum frequency of BiBaK is no more than 120 cm$^{-1}$, which is much lower than those of the other three HHs. Moreover, we observe obvious hybridization of acoustic branches and low-frequency optical branches in BiBaK, which means that phonon-phonon scatterings are more likely to occur than the other three systems. Figure 3(a) plots the phonon group velocities ($v$) of BiBaK as a function of frequency. For the TA and LA branches, the calculated mean values are 1433 m/s and 2374 m/s, respectively. These results are comparable to those of good thermoelectric materials with lower lattice thermal conductivity, such as Bi$_2$Te$_3$ ($\kappa_l$=1.2 W/mK, $v_{TA}$=1630 m/s, $v_{LA}$=2650 m/s) [34] and PbTe ($\kappa_l$=2.1 W/mK, $v_{TA}$=1610 m/s, $v_{LA}$=3596 m/s) [35,36]. In contrast, the phonon group velocities of other three HHs in Fig. 2 are much higher, as indicated by their strong phonon dispersion relations. Furthermore, the calculated relaxation time of most phonon modes of BiBaK is in the range of 1~100 ps, which is comparable to that of Bi$_2$Te$_3$ [37]. All these findings suggest the BiBaK should have very small lattice thermal conductivity. Indeed, we see from Fig. 3(b) that the $\kappa_l$ is 1.82 at 300 K, which is very close to that calculated previously [18] and also confirms the reliability of our computational approach. As known, the $\kappa_l$ decreases with temperature and the value is only 0.60 at 900 K.

To figure out the physical origin of the intrinsically low $\kappa_l$ of the BiBaK compound, we show in Fig. 3(c) and Fig. 3(d) the projected phonon density of states (PDOS) and the cumulative lattice thermal conductivity, respectively. In the low-frequency region, we see that the Bi atom dominates the PDOS and the $\kappa_{cumu}$ increases quickly with the frequency. The rise becomes slowly in the medium-frequency region where the contribution from Ba atom is nearly the same as that of Bi atom, and the weak bonding between Bi and Ba atoms as well as the heavier atomic mass can result in lower sound velocity and Young's modulus [6,38,39]. Beyond the frequency gap in the range of 80~100 cm$^{-1}$, we find that $\kappa_{cumu}$ keeps almost unchanged where the K atom governs the PDOS, which suggests that it can lead to much stronger anharmonic scattering between phonons. To go further, we plot in Figure 4(a) the temperature dependence of the atomic displacement parameters (ADP) of BiBaK. We see that the ADP of K atom is obviously larger than those of Bi and Ba atoms, and their differences become more and more pronounced at elevated temperature. It should be mentioned that the cage-like materials such as skutterudites and clathrates have been suggested as good thermoelectric systems [40−43] due to the well-known concept of "phonon-glass and electron-crystal" [40,44,45]. In such kind of structures, the multiple fillings of the portions in the host nanocages bring non-overlapping phonon vibrations so that the lattice thermal conductivity can be efficiently reduced [46]. The obviously larger ADP of K atom compared with those of Bi and Ba atoms is reminiscent of guest atoms "rattling" in the host cages. Indeed, we see from Fig. 4(b) that around each K atom, there is a hexadecahedron cage consisting of 4 Bi atoms and 6 Ba atoms in the crystal structure of BiBaK. Hence, it is reasonable to conclude that the heavy Bi and Ba atoms give the most contribution to $\kappa_l$. Meanwhile, the rattling of the guest K atoms inside the cage induces obvious phonon scattering and consequently reduces the $\kappa_l$.

Figure 5 plots the electronic band structure of BiBaK, where we find an indirect band gap of ~0.75 eV with the conduction band minimum (CBM) and the valence band maximum (VBM) located at the $\Gamma$ and $X$ points, respectively. Note that both the CBM and VBM are doubly degenerated. The energy band near the CBM forms a steep valley while that around VBM is relatively flat, which suggests quite different effective mass for electrons and holes. It was previously demonstrated that high carrier mobility may stem from a steep valley at the $\Gamma$ point, where the valley-valley or peak-valley scatterings are inhibited [47]. It is thus reasonable to expect that the *n*-type BiBaK may have favorable electronic transport properties, as discussed in the following.

Before evaluating the transport coefficients, it is imperative to figure out the carrier relaxation time. As the major scattering mechanism is acoustic phonons, we adopt the deformation potential (DP) theory, where the relaxation time is given by $\tau = 2\sqrt{2\pi} C \hbar^4 / 3(k_B T m_{dos}^*)^{3/2} E^2$ [48] with $C$, $m_{dos}^*$ and $E$ are the elastic module, the DOS effective mass, and the DP constant, respectively. Table 1 summarizes the room temperature relaxation time of BiBaK for electrons and holes. We see that the relaxation time of electrons is obviously larger than that of holes, which can be attributed to their considerably different DOS effective mass, as also evidenced by the dispersion relations near the CBM and VBM (Fig. 5). Using the semi-classical Boltzmann theory and inserting the calculated carrier relaxation time, we can obtain the electronic transport coefficients of BiBaK ($S$, $\sigma$, and $\kappa_e$) at various temperatures and concentrations. In particular, our calculations find that the system exhibits higher power factor ($S^2\sigma$) of $2.38 \times 10^{-3}$ W/mK$^2$ and $7.28 \times 10^{-4}$ W/mK$^2$ at 900 K for the *n*- and *p*-type carriers, respectively. In combination with the phonon transport coefficient ($\kappa_l$) discussed above, we can now evaluate the thermoelectric performance of BiBaK. Figure 6(a) shows the calculated *ZT* values as a function of carrier concentration at two typical temperatures of 300 and 900 K. In both cases, it is clear that the *ZT* can be maximized by optimizing the carrier concentration. If we focus on the high temperature, we see that the *p*-type system exhibits a peak *ZT* of ~0.9 at the concentration of $4.3 \times 10^{20}$ cm$^{-3}$. For the *n*-type system, however, the *ZT* value can be significantly optimized to 1.9 at much lower concentration of $4.2 \times 10^{18}$ cm$^{-3}$, which outperforms many good thermoelectric materials. In Fig. 6(b), we plot the *ZT* value as a function of temperature from 300 to 900 K. We see that the *ZT* of *n*-type BiBaK increases almost linearly with the temperature, and is obviously higher than that of *p*-type system in the whole temperature considered. Much efforts should be thus devoted to enhance the thermoelectric performance of *p*-type BiBaK so that comparable efficiencies could be realized in the fabrication of both *n*- and *p*-legs of thermoelectric modules.

## 4. Summary

We have performed a theoretical study on the structural, electronic, phonon, and thermoelectric transport properties of the HH compound BiBaK. Among three possible atomic configurations, we identify the most stable structure with an indirect band gap of 0.75 eV by adopting the hybrid functional and considering the SOC. Due to the rattling of K atoms inside the hexadecahedron cage formed by Bi and Ba atoms, the BiBaK exhibits unprecedentedly low lattice thermal conductivity, which consequently leads to a high *ZT* value of 1.9 at 900 K with a realistic *n*-type carrier concentration of $4.2 \times 10^{18}$ cm$^{-3}$. It would be interesting to investigate in the future work if favorable thermoelectric performance could be also found in other similar HHs such as SnBaSr, PdSrTe, and TeAgLi, which are assumed to have rather small lattice thermal conductivity at room temperature [18].

**Acknowledgements**

We thank financial support from the National Natural Science Foundation (Grant Nos. 51772220 and 11574236). The numerical calculations in this work have been done on the platform in the Supercomputing Center of Wuhan University.

**Table 1** The elastic constant $C$, the deformation potential constant $E$, the DOS effective mass $m^*_{dos}$ and the carrier relaxation time $\tau$ of BiBaK at room temperature.

| Carrier type | $C$ (eV/Å$^3$) | $E$ (eV) | $m^*_{dos}$ ($m_e$) | $\tau$ (s) |
|---|---|---|---|---|
| Electron | 0.188 | −5.75 | 0.257 | $2.43 \times 10^{-13}$ |
| Hole | 0.188 | −5.18 | 1.49 | $2.15 \times 10^{-14}$ |

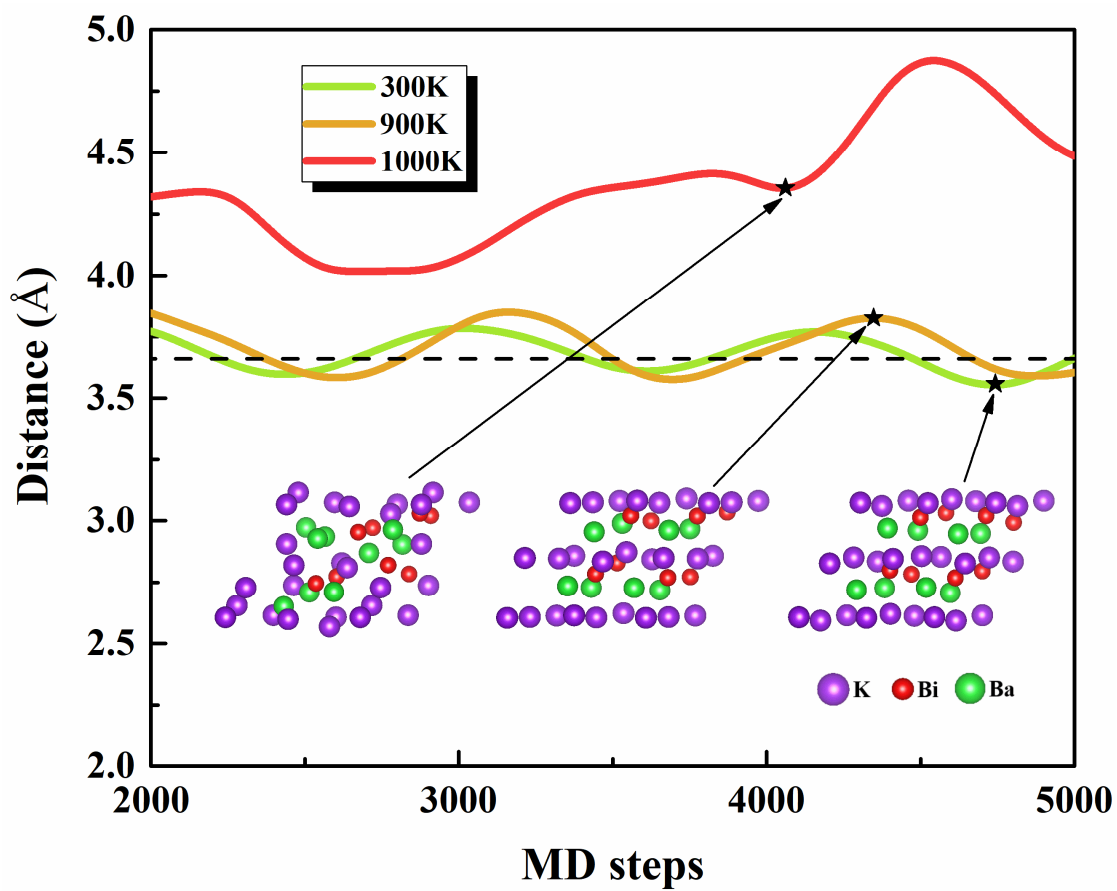

**Figure 1.** The AIMD results of the nearest Bi-Ba distance for BiBaK at 300 K, 900 K, and 1000 K. The three insets correspond to the crystal structures at 4050, 4346 and 4738 MD step, respectively. The dash line indicates the equilibrium Bi-Ba distance.

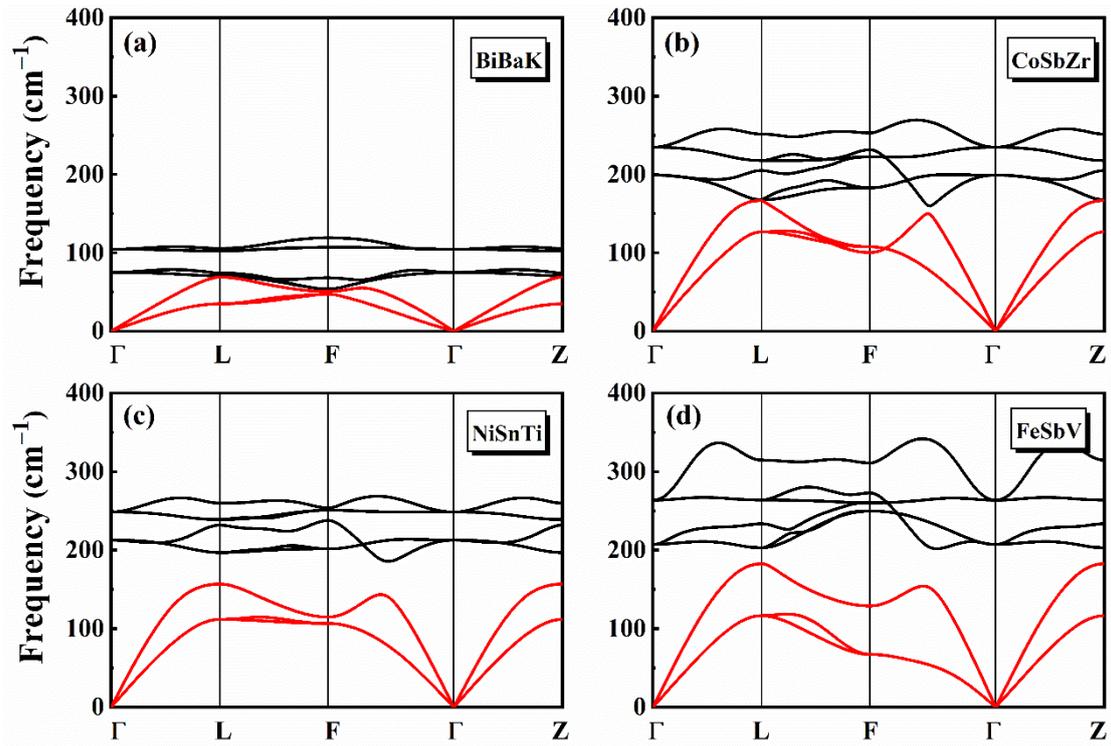

**Figure 2.** The phonon dispersion relations of (a) BiBaK, (b) CoSbZr, (c) NiSnTi, and (d) FeSbV, where the red lines indicate acoustic branches and black lines for optical ones.

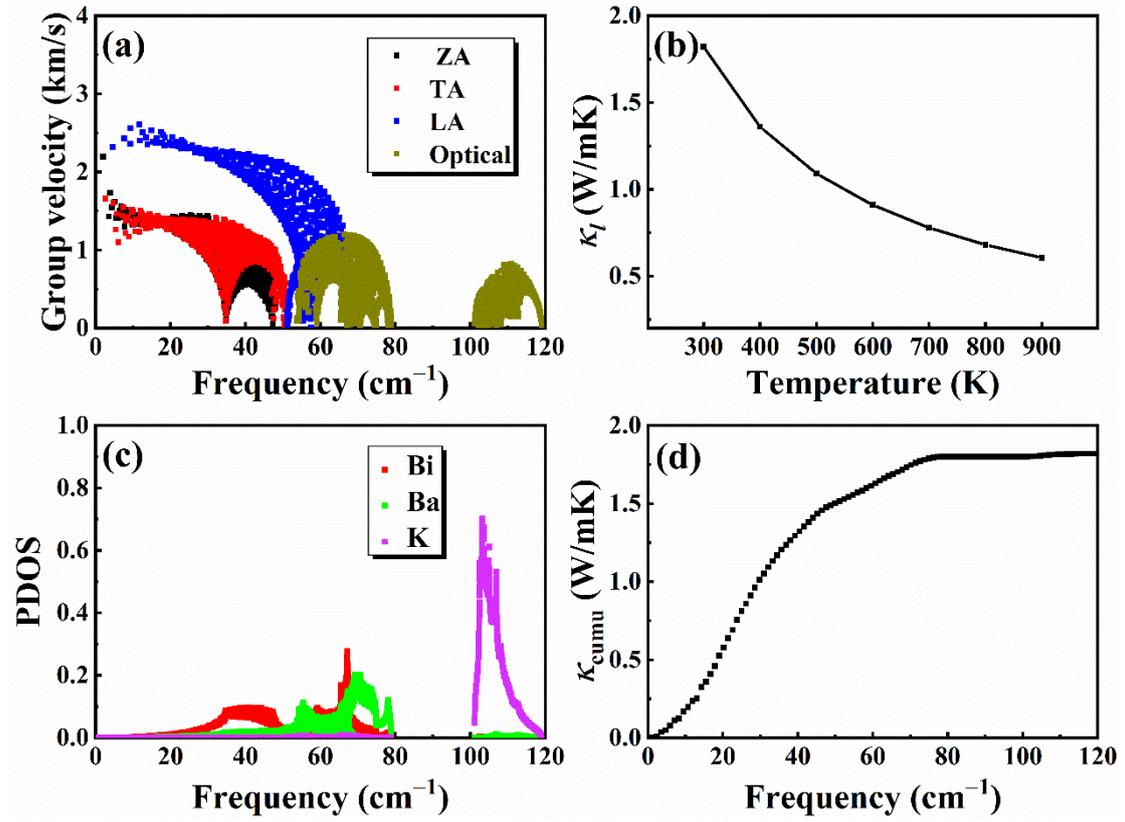

**Figure 3.** (a) The group velocity of different phonon modes in the BiBaK plotted as a function of frequency. (b) The temperature dependence of the lattice thermal conductivity. (c) The projected phonon density of states. (d) The cumulative lattice thermal conductivity at room temperature plotted as function of phonon frequency.

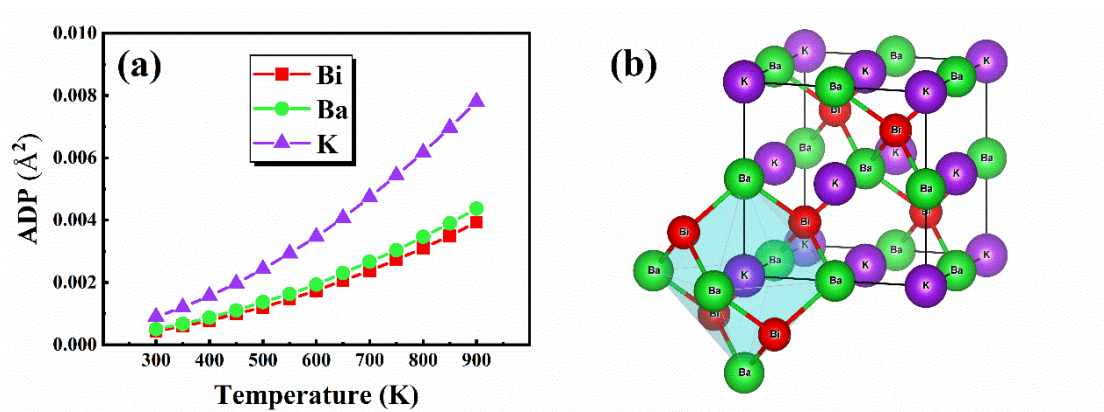

**Figure 4.** (a) The calculated atomic displacement parameters of BiBaK as a function of temperature. (b) The BiBaK exhibits a cage-like structure with the guest K atom rattles inside the hexadecahedron host framework consisting of 4 Bi and 6 Ba atoms.

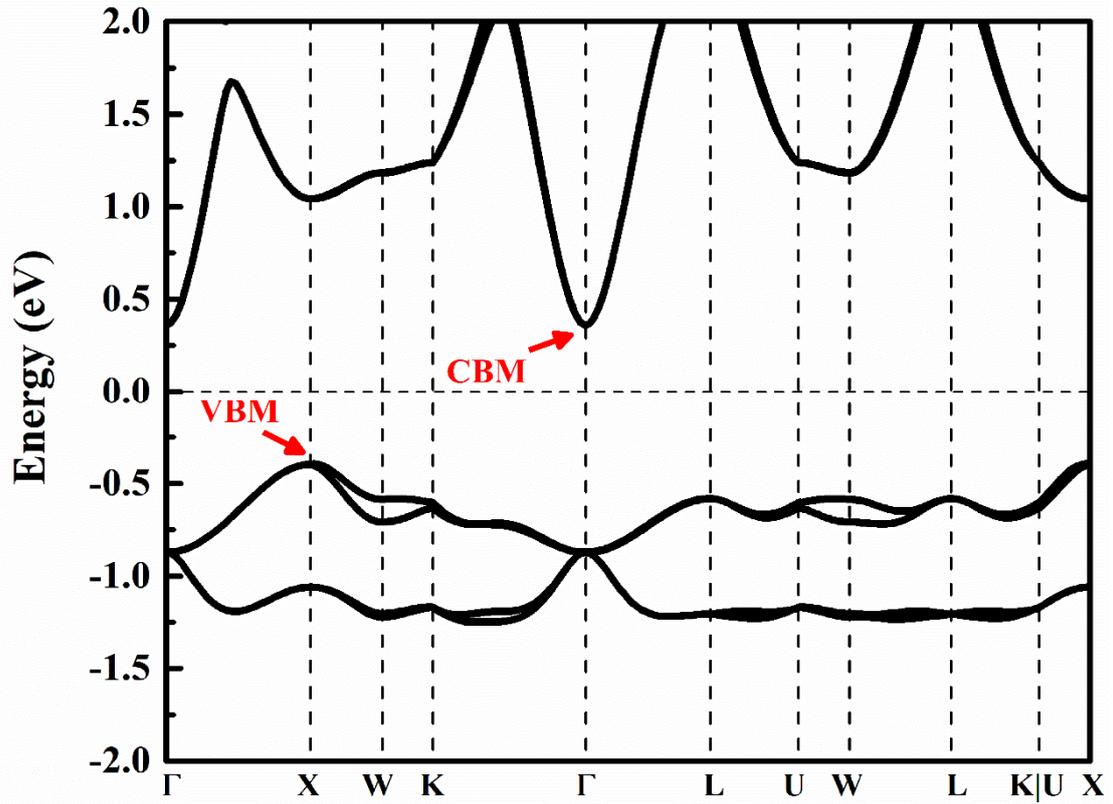

**Figure 5.** The calculated energy band structure of BiBaK. The Fermi level is at 0 eV.

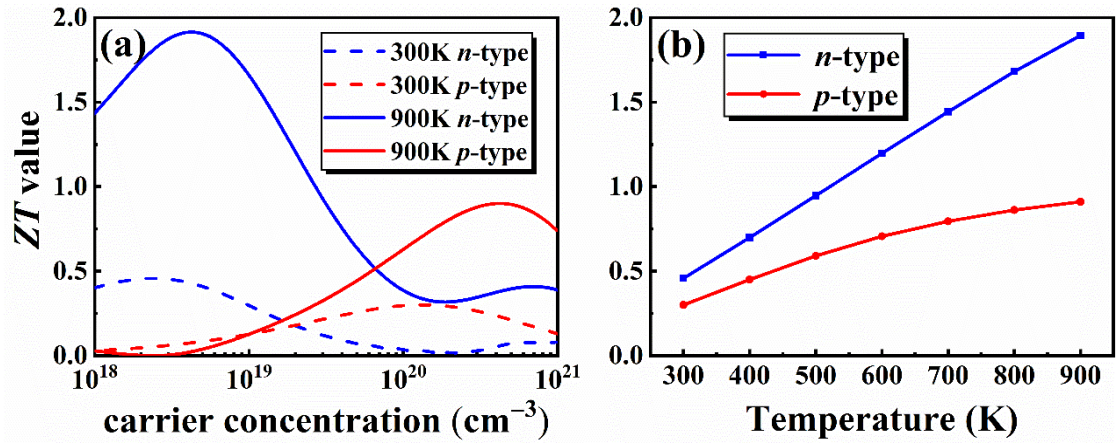

**Figure 6.** (a) The calculated *ZT* values of BiBaK as a function of carrier concentration at 300 K and 900 K. (b) Temperature dependent *ZT* values for both *n*- and *p*-type systems.